\documentclass[11pt]{article}

\usepackage{SIGACTNews}
\usepackage{chicagor}
\usepackage{amsmath}
\usepackage{amssymb}
\usepackage{proof}

\newcommand{\thmcolon}{\hspace{-1ex}{\bf .}}

\newtheorem{THEOREM}{Theorem}%
\newtheorem{LEMMA}[THEOREM]{Lemma}
\newtheorem{COROLLARY}[THEOREM]{Corollary}
\newtheorem{PROPOSITION}[THEOREM]{Proposition}
\newtheorem{DEFINITION}[THEOREM]{Definition}
\newtheorem{CLAIM}[THEOREM]{Claim}
\newtheorem{EXAMPLE}[THEOREM]{Example}
\newtheorem{REMARK}[THEOREM]{Remark}

                        {\end{THEOREM}}
                      {\end{LEMMA}}
                          {\end{COROLLARY}}
\newenvironment{proposition}{\begin{PROPOSITION} \thmcolon  }%
                            {\end{PROPOSITION}}
                            { \end{DEFINITION}}
                            {\end{CLAIM}}

                            { \wbox\end{EXAMPLE}}
\newenvironment{example*}{\begin{EXAMPLE} \thmcolon  \rm}%
                            {\end{EXAMPLE}}
                            { \wbox\end{REMARK}}
\newenvironment{remark*}{\begin{REMARK} \thmcolon  \rm}%
                            {\end{REMARK}}

\def\squareforqed{\hbox{\rlap{$\sqcap$}$\sqcup$}}
\def\wbox{\ifmmode\squareforqed\else{\unskip\nobreak\hfil
\penalty50\hskip1em\null\nobreak\hfil\squareforqed
\parfillskip=0pt\finalhyphendemerits=0\endgraf}\fi}

\newcommand{\Rule}[2]{          %
  \begin{array}{c}
  #1 \\\hline
  #2
  \end{array}}

\newenvironment{prog}{\begin{array}[t]{@{}l@{}}}{\end{array}}

\usepackage{url}

\newcommand{\splitcolumn}{\vskip 2em\hrule\vskip 3em}

\newcommand{\sat}{\models}
\newcommand{\rimp}{\Rightarrow}
\newcommand{\riff}{\Leftrightarrow}
\renewcommand{\phi}{\varphi}
\newcommand{\<}{\langle}
\renewcommand{\>}{\rangle}

\newcommand{\intension}[1]{[\![ #1 ]\!]}

\newcommand{\red}[1]{\stackrel{#1}{\longrightarrow}}
\newcommand{\R}{\mathcal{R}}
\newcommand{\cA}{\mathcal{A}}

\newcommand{\true}{\mbox{\textbf{true}}}
\newcommand{\false}{\mbox{\textbf{false}}}
\newcommand{\truep}{\mathit{true}}

\newcommand{\RuleSide}[3]{
  \begin{array}{c}
  #1 \\\hline #2
  \end{array}
  {~~#3}}

\newcommand{\collapse}{\bowtie}
\newcommand{\dcup}{\uplus}
\newcommand{\mi}[1]{\mathit{#1}}

\newcommand{\COMMENTOUT}[1]{}

\title{SIGACT News Logic Column 12\\[2ex]
\textbf{Logical Verification and Equational Verification}\footnote{\copyright{} Riccardo Pucella, 2005.}} 
\author{Riccardo Pucella\\
Cornell University\\
Ithaca, NY 14853 USA\\
riccardo@cs.cornell.edu}
\date{}

\begin{document}

\SIGACTmaketitle

\noindent\textbf{Note from the Editor:} There were some errors in the
last article of the Logic Column. Thanks to Claudia Zepeda for
spotting them. They have been corrected in the online version of the
article, available from the CORR archive at
\url{http://arxiv.org/abs/cs.LO/0502031}. All articles published in
this column are archived at CORR; the following URL will return them
all:  \url{http://arxiv.org/find/grp_cs/1/ti:+AND+logic+column}. 

I am always looking for contributions. If you have any suggestion
concerning the content of the Logic Column, or even better, if you
would like to contribute by writing a survey or tutorial on your own
work or topic related to your area of interest, feel free to get in
touch with me.

\splitcolumn

\noindent One of the main uses of logic in computer science is in verification,
that is, specifying properties of systems (I use the term generally),
and proving that systems satisfy those properties. A quick survey of
the verification literature reveals two popular approaches.  

The first approach relies on a logic in which to express properties of
systems, and on the definition of a satisfaction relation to prescribe
when a formula of the logic is true of a system. One can then develop
techniques such as model checking or theorem proving for establishing
that a property is true of a system. Let me call this approach
\emph{logical verification}. 

The second approach is based on the following observation: to verify
that a system has a particular property, it suffices to show the
system equivalent to another system that obviously has the desired
property. The technical meat of such an approach consists in defining
suitable notions of equivalence, establishing that equivalence
preserves the properties of interest, and developing techniques for
proving the equivalence of two systems. Such techniques generally
involve manipulating equations involving equivalence between
systems. Accordingly, let me call this approach
\emph{equational verification}. 

The distinction between logical verification and equational
verification is not a new one. One finds early discussions of a
similar distinction in the distributed computing literature. The
transition axiom method of Lamport \citeyear{r:lamport83}, essentially
a form of equational verification, was admittedly developed to
compensate for perceived insufficiencies in logical verification based
on temporal logic (when used to specify properties of distributed
systems). I find it interesting that the distinction occurs in so many
places in the literature, in many different fields. The purpose of
this article is to illustrate the distinction and examine the two
approaches in a number of simple settings where their relationship
is particularly easy to describe, and hopefully to wet your appetite
and let you investigate other areas where the distinction between
logical verification and equational verification arises.

\section*{Processes}

Process calculi emerged from the work of Hoare \citeyear{r:hoare85} and
Milner \citeyear{r:milner80} on models of concurrency, and are meant to
model systems made up of processes communicating by exchanging values
across channels. They allow for the dynamic creation and removal of
processes, allowing the modeling of dynamic systems.  A typical
process calculus is CCS \cite{r:milner80,r:milner89}, which is the
foundation of a number of more involved calculi. 

CCS provides a minimalist syntax for writing processes.  Processes
perform actions, which can be of three forms: the sending of a message
over channel $x$ (written $\overline{x}$), the receiving of a message
over channel $x$ (written $x$), and internal actions (written $\tau$),
the details of which are unobservable.\footnote{In the literature, the
actions of CCS are often given a much more abstract interpretation, as
simply names and co-names. The send/receive interpretation is useful
for being easy to grasp.}  Send and receive actions are called
\emph{synchronization} actions, since communication occurs when the
corresponding processes synchronize. Let $\alpha$ stand for actions,
including the internal action $\tau$, while  $\lambda$ is reserved for
synchronization actions. The syntax of CCS processes is given by the
following grammar: 
\[
P,Q ::= \alpha_1.P_i +\dots +\alpha_n.P_n \mid P_1|P_2 \mid \nu x.P.
\]
We write $0$ for the empty sum (when $n=0$). The process $0$
represents the process that does nothing and simply terminates. A
process of the form $\lambda.P$ awaits to synchronize with a process
of the form $\overline{\lambda}.Q$, after which the processes continue
as process $P$ and $Q$ respectively. A generalization of such
processes is $\alpha_1.P_1+\dots +\alpha_n.P_n$, which
nondeterministically synchronizes via a single $\alpha_i$. 
To combine processes, the parallel composition $P_1 | P_2$ is
used. The difference between sum and parallel composition is that a
sum offers a choice, so only one of the summands can
synchronize and proceed, while a parallel composition allows all its
component processes to proceed. 
The process  $\nu x.P$ defines a local channel name $x$ to
be used within process $P$. This name is guaranteed to be unique to
$P$ (possibly through consistent renaming).

As an example, consider the process $(x.y.0 + \overline{x}.z.0) |
\overline{x}.0 | \overline{y}.0$. Intuitively, it consists of three
processes running in parallel: the first offers of choice of either
receiving over channel $x$, or sending over channel $x$, the second
sends over channel $x$, and the third sends over channel
$y$. Depending on which choice the first process performs (as we will
see, this depends on the actions the other process can perform), it
can continue in one of two ways: if it chooses to receive on channel
$x$ (i.e., the $x.y.0$ summand is chosen), it can then receive on
channel $y$, while if it chooses to send on $x$ (i.e., the
$\overline{x}.z.0$ summand is chosen), it can then receive on channel
$z$. 

To represent the execution of a process, we define the
notion of a transition. Intuitively, the transition relation tells us
how to perform one step of execution of the process. Note that since
there can be many ways in which a process executes, the transition is
fundamentally nondeterministic. The transition of a process $P$ into a
process $Q$ by performing an action $\alpha$ is indicated
$P\red{\alpha} Q$. The action $\alpha$ is the observation of the
transition. 
The transition relation is defined by the following inference rules: 
\[\begin{array}{ccc}
\multicolumn{3}{c}{
\RuleSide{}
            {\alpha_1.P_1+\dots+\alpha_n.P_n\red{\alpha_j}P_j}
            {\mbox{for $j\in 1..n$}}
}\\[5ex]
\RuleSide{P\red{\alpha}P'}
            {\nu x.P\red{\alpha}\nu x.P'}
            {\mbox{if $\alpha\not\in\{x,\overline{x}\}$}} 
& \qquad & 
\Rule{P\red{\alpha}P'}
        {P|Q\red{\alpha}P'|Q} 
\\[5ex]
\Rule{P\red{\lambda} P'\quad Q\red{\overline{\lambda}}Q'}
        {P|Q\red{\tau} P'|Q'}
& & 
\Rule{Q\red{\alpha}Q'}
        {P|Q\red{\alpha}P|Q'.} 
\end{array}\]

For example, consider the transitions of the example process above,
$(x.y.0 + \overline{x}.z.0) | \overline{x}.0 | \overline{y}.0$. A
possible first transition (the first step of the execution, if you
wish), can be derived as follows:
\[
  \infer{(x.y.0 + \overline{x}.z.0) | \overline{x}.0 | \overline{y}.0\red{\tau}y.0 | 0 | \overline{y}.0.}
        {\infer{(x.y.0 + \overline{x}.z.0) | \overline{x}.0\red{\tau} y.0 | 0}
               {x.y.0 + \overline{x}.z.0\red{x} y.0 & 
                \overline{x}.0 \red{\overline{x}} 0}}
\]
That is, the process reduces to $y.0 | 0 | \overline{y}.0$ in one step
that does not provide outside information, since it appears as an
internal action. (The $0$ can be removed from the resulting process,
as it does not contribute further to the execution of the process,
although there is an argument that goes here and that I am shamelessly
sweeping under the rug.). The resulting process $y.0 | \overline{y}.0$
can then perform a further transition, derived as follows:
\[
  \infer{y.0 | \overline{y}.0 \red{\tau} 0|0.}
        {y.0 \red{y} 0 & 
         \overline{y}.0 \red{\overline{y}} 0}
\]
In summary, a possible sequence of transitions for the original
process is the two-step sequence \[(x.y.0 + \overline{x}.z.0) |
\overline{x}.0 | \overline{y}.0\red{\tau}y.0 | \overline{y}.0
\red{\tau} 0.\]

A logic for reasoning about CCS processes was introduced by Hennessy
and Milner \citeyear{r:hennessy85}, called simply Hennessy-Milner
logic, or HML. The syntax of HML formulas is given by the following
grammar:
\[ 
\phi ::= \truep \mid \neg\phi \mid \phi_1\land\phi_2 \mid
[\alpha]\phi.
\]
Formulas describe properties of processes. The formula $\truep$
represents the formula which is true of every process. The formula
$\neg\phi$ is the negation of $\phi$, while $\phi_1\land\phi_2$ is the
conjunction of $\phi_1$ and $\phi_2$. The formula $[\alpha]\phi$,
where $\alpha$ is an action (possibly $\tau$), intuitively says that
for all ways that a process can perform  action $\alpha$, it
transitions to a process for which $\phi$ is true. We define the
disjunction $\phi_1\lor\phi_2$ as an abbreviation for
$\neg(\neg\phi_1\land\neg\phi_2)$ and the implication
$\phi_1\rimp\phi_2$ as an abbreviation for $\neg\phi_1\lor\phi_2$. 

Following the intuitions outlined above, we can formally define what
it means for a formula $\phi$ to be true for process $P$, written
$P\sat\phi$, by induction on the structure of $\phi$:
\begin{itemize}
\item[] $P\sat \truep$  always
\item[] $P\sat\neg\phi$ if $P\not\sat\phi$
\item[] $P\sat\phi_1\land\phi_2$ if $P\sat\phi_1$ and
$P\sat\phi_2$
\item[] $P\sat [\alpha]\phi$ if for all $Q$ such that
$P\red{\alpha}Q$, $Q\sat\phi$.  
\end{itemize}

Instead of using a logic such as HML, a popular way of reasoning about
processes is to consider a definition of equivalence between processes,
and reason equationally about the equivalence of a process with an
``ideal'' process that represents the correct desired behavior. One
standard notion of equivalence between processes is to take two
processes as equivalent if they are indistinguishable from the point
of view of an external observer interacting with the processes. A
particular formalization of such an equivalence is \emph{strong
bisimilarity} \cite{r:milner80}. A strong bisimulation is a relation
$\R $ such that whenever $(P,Q)\in\R$, we have:
\begin{itemize}
\item If $P\red{\alpha}P'$, then there exists
$Q'$ such that $Q\red{\alpha}Q'$ and $(P',Q')\in\R$;
\item If $Q\red{\alpha}Q'$, then there exists $P'$ such that $P\red{\alpha}P'$
and $(P',Q')\in\R$.
\end{itemize}
We say $P$ and $Q$ are \emph{strongly bisimilar}, written $P\sim Q$,
if there exists a strong bisimulation $\R $ such that $(P,Q)\in\R$. In
other words, if $P$ and $Q$ are strongly bisimilar, then whatever
transition $P$ can take, $Q$ can match it with one of his own that
results in processes that are themselves strongly bisimilar, and vice
versa. It is easy to check that $\sim$ is an equivalence relation.

The following result of Hennessy and Milner \citeyear{r:hennessy85}
highlights the deep relationship between logical verification using HML
and equational verification  using strong bisimilarity.
\begin{proposition}\label{p:hml-equivalence}
The following are equivalent:
\begin{enumerate}
\item[(a)] $P\sim Q$;
\item[(b)] For all $\phi$, $P\sat\phi$ if and only if $Q\sat\phi$. 
\end{enumerate}
\end{proposition}

Based on this observation, we can associate with a process $Q$ (when
viewed as a specification) the set of HML formulas that $Q$
satisfies, $\intension{Q}\triangleq\{\phi\mid Q\sat\phi\}$. A simple
recasting of Proposition~\ref{p:hml-equivalence} gives:
\begin{proposition}
The following are equivalent:
\begin{enumerate}
\item[(a)] $P\sim Q$;
\item[(b)] For all $\phi$, $P\sat\phi$ if and only if $\phi\in\intension{Q}$. 
\end{enumerate}
\end{proposition}
Thus, a process $P$ is strongly bisimilar to a ``specification
process'' $Q$ if it satisfies exactly the formulas $\intension{Q}$
associated with $Q$. In this sense, we can understand strong
bisimilarity as checking that a particular class of formulas holds of
a process.

Similar results can be obtained for different notions of
equivalence. Strong bisimilarity is a very fine equivalence relation;
not many processes end up being equivalent. More worryingly, strong
bisimilarity does not handle internal actions very well. Intuitively,
process equivalence should really only involve observable actions. Two
processes that only perform internal actions should be considered
equivalent. For instance, the processes $\tau.\tau.0$ and $\tau.0$
should really be considered equivalent, as they really do nothing
after performing some internal (and hence really unobservable)
actions. Unfortunately, it is easy to check that these two processes
are not strongly bisimilar. To address this situation, a weaker notion
of equivalence, weak bisimilarity, is often used in practice. It is
easy to extend HML to deal with this form of equivalence, by
introducing suitable modal operators. In a similar way, we can extend
HML to express recursive properties; this leads quickly the
propositional modal $\mu$-calculus
\cite{r:kozen83}. For much more along those lines, see the excellent
monograph of Stirling \citeyear{r:stirling01}. Furthermore, we can
play this game of logical and equational verification for
processes in the context of the $\pi$ calculus, which extends CCS with
the capability to send values over channels \cite{r:milner99}. An
extension of HML capturing some forms of equivalence for the $\pi$
calculus is given by Milner, Parrow, and Walker
\citeyear{r:milner93a}.

\section*{Programs}

A similar, but slightly more complicated picture, arises for reasoning
about programs. To keep things as clear as possible, let me consider a
very simple class of programs, regular programs. Start with a set
$\cA$ of \emph{primitive programs}. We use $a$ to range over primitive
programs. Primitive programs are abstract operations we want our
programs to perform. The syntax of regular programs is given by the
following grammar:
\[ 
\alpha,\beta ::= a \mid \alpha_1;\alpha_2 \mid \alpha_1+\alpha_2 \mid \alpha^*.
\]
Thus, a primitive program $a\in\cA$ is a regular program. The program
$\alpha_1;\alpha_2$ represents the sequencing of programs $\alpha_1$ and
$\alpha_2$, while the program $\alpha_1+\alpha_2$ represents a
nondeterministic choice between programs $\alpha_1$ and $\alpha_2$. The
program $\alpha^*$ represent the finite iteration of the program
$\alpha$ for a nondeterministic number of iterations (possibly
none). Of course, the name ``regular programs'' comes from the fact
that we can view such programs as regular expressions.

It is well known how to give a semantics to regular programs: we map a
program to a relation between initial states and final states.  An
interpretation for the primitive programs is a map $\sigma$ that
associates to each primitive program $a$ a binary relation $\sigma(a)$
on the set of states. Intuitively, $(s_1,s_2)\in\sigma(a)$ if
executing the primitive program $a$ in state $s_1$ leads to state
$s_2$. We give a semantics to arbitrary programs by extending $\sigma$
inductively to all programs:
\begin{align*}
\sigma(\alpha_1;\alpha_2) & \triangleq \sigma(\alpha_1) \circ \sigma(\alpha_2)\\
\sigma(\alpha_1+\alpha_2) & \triangleq \sigma(\alpha_1)\cup\sigma(\alpha_2)\\
\sigma(\alpha^*) & \triangleq \bigcup_{n\geq 0} \sigma(\alpha)^n.
\end{align*}
For $R$ and $S$ binary relations, we write $R \circ S$ for the
relation $\{(u,v) \mid \exists w. (u,w)\in R, (w,v)\in S\}$, and $R^n$
is defined inductively by taking $R^0$ to be the identity relation,
and $R^{n+1}$ to be $R^n \circ R$. The map $\sigma$ gives us what is
commonly called the input-output semantics of programs.

To reason about properties of those programs, we consider a logic
called propositional dynamic logic, or PDL \cite{r:harel00}.  We start
with a set $\Phi_0$ of primitive propositions representing basic facts
about states.  We use $p$ to range over primitive propositions. The
syntax of PDL formulas is given by the following grammar:
\[
\phi ::= p \mid \neg\phi \mid \phi_1\land\phi_2 \mid [\alpha]\phi.
\]
Thus, a primitive proposition $p\in\Phi_0$ is a formula.  The formulas
$\neg\phi$ and $\phi_1\land\phi_2$ have their usual reading, while the
formula $[\alpha]\phi$, where $\alpha$ is a regular program, reads
``all halting executions of program $\alpha$ result in a state
satisfying $\phi$''. As in the last section, we define
$\phi_1\lor\phi_2$ as an abbreviation for
$\neg(\neg\phi_1\land\neg\phi_2)$ and $\phi_1\rimp\phi_2$ as an
abbreviation for $\neg\phi_1\lor\phi_2$. Furthermore, we define
$\phi_1\riff\phi_2$ as an abbreviation for
$(\phi_1\rimp\phi_2)\land(\phi_2\rimp\phi_1)$. We write
$\<\alpha\>\phi$ as an abbreviation for $\neg[\alpha]\neg\phi$;
$\<\alpha\>\phi$ reads ``at least one halting execution of $\alpha$
results in a state satisfying $\phi$''.

The semantics of PDL is given using Kripke structures
\cite{r:kripke63}. Essentially, a Kripke structure is a set of states; think of all 
the states a program could be in. Following the formalization above,
programs are interpreted as a relation between initial states and
final states, and at every state we have an interpretation function
telling us what primitive propositions are true at that state. General
formulas will express properties of moving through that state
space. Formally, a Kripke structure $M$ is a tuple $(S,\pi,\sigma)$
where $S$ is a set of states, $\pi$ is an interpretation function
assigning a truth value to each primitive proposition $p$ at each
state $s$ (i.e., $\pi(s)(p)\in\{\true,\false\}$) and $\sigma$ is an
interpretation for the primitive programs, as defined earlier.  We
define what it means for a PDL formula $\phi$ to be true in state $s$
of $M$, written $(M,s)\sat\phi$, by induction on the structure of
$\phi$:
\begin{itemize}
\item[] $(M,s)\sat p$  if $\pi(s)(p)=\true$
\item[] $(M,s)\sat \neg\phi$ if $(M,s)\not\sat\phi$
\item[] $(M,s)\sat \phi_1\land\phi_2$ if $(M,s)\sat\phi_1$ and
$(M,s)\sat\phi_2$
\item[] $(M,s)\sat[\alpha]\phi$ if for all
$s'$ such that $(s,s')\in\sigma(\alpha)$, $(M,s')\sat\phi$.
\end{itemize}
Thus, a formula $[\alpha]\phi$ is true at a state $s$ if for
all states $s'$ that can be reached by executing the program $\alpha$
at state $s$, $\phi$ holds. We can verify that $\<\alpha\>\phi$ holds
at a state $s$ if and only if there is at least one state that is
reachable by program $\alpha$ from state $s$ such that $\phi$ holds in 
the state, hence justifying our intuitive reading of $\<\alpha\>\phi$.
If a formula $\phi$ is true at all the states of a model $M$, we say
that $\phi$ is valid in $M$ and write $M\sat\phi$. If a formula $\phi$ 
is valid in all models, we say $\phi$ is valid, and write $\sat\phi$. 

An alternative approach for reasoning about regular programs is to
give a direct definition of equivalence between programs, and use
equational logic to reason about equivalence of programs. For regular
programs, a popular notion of equivalence is obtained by taking two
programs to be equivalent if they denote the same input-output
relation on states. This kind of equivalence is captured by the theory
of Kleene algebras \cite{r:conway71}. One presentation of this theory
is the following axiomatization given by Kozen  \citeyear{r:kozen94}:
\[\begin{array}{lclcl}
x+(y+z) = (x+y)+z & \qquad & x+y=y+x & \qquad & 1+x;x^* \le x^*\\
x+0=x & & x+x=x & & 1+x^*;x \le x^*\\
x;(y;z)=(x;y);z & & 1;x=x;1=x & & b+a;x \le x \rimp a^*;b \le x\\
x;(y+z) = x;y + x;z & & (x+y);z = x;z + y;z & & b+x;a \le x \rimp b;a^* \le x\\
0;x = x;0 = 0, & &  & & 
\end{array}\]
where $0$ and $1$ are constants representing respectively the empty
program (with no executions) and the identity program (that does not
change the state), and $x\le y$ if and only if $x+y=y$. 

By general considerations of equational logic, the axioms of Kleene
algebra along with the usual axioms for equality, instantiation, as
well as rules for introduction and elimination of implications,
constitutes a complete deductive system for reasoning about
implications in the theory of Kleene algebras
\cite{r:selman72}.\footnote{More precisely, it constitutes a complete
deductive system for the so-called universal Horn theory of Kleene
algebras.}

The fundamental relationship between equational verification using
Kleene algebras and logical verification using PDL is given by the
following result, which partly follows from the fact that the
relational semantics of programs given above forms a Kleene algebra:
\begin{proposition}\label{p:pdl-equivalence}
The following are equivalent:
\begin{enumerate}
\item[(a)] $\alpha=\beta$;
\item[(b)] For all $\phi$, $\sat\<\alpha\>\phi\Leftrightarrow \<\beta\>\phi$.
\end{enumerate}
\end{proposition}
One distinct advantage with reasoning equationally is that deciding
$\alpha=\beta$ is PSPACE-complete \cite{r:stockmeyer73}, while
deciding $\sat\phi$ in PDL is EXPTIME-complete
\cite{r:fischer79,r:pratt78}.

Is there a way to view a program $\beta$ as a specification against
which we can check another program $\alpha$? Recall what we did for
processes in the previous section: a process $P$ was equivalent to a
``specification process'' $Q$ if $P$ satisfied exactly the HML
formulas associated with the specification process $Q$. Can something
similar be done for programs? The answer is yes, and here is a natural
way to do it. It relies intrinsically on the relationship between
regular expressions and finite automata, which are almost Kripke
structures.  By associating with a program $\alpha$ a Kripke structure
representing that program, we can associate with a ``specification
program'' $\beta$ the set of PDL formulas that are valid in the Kripke
structure representing $\beta$. One would then hope that $\alpha$ and
$\beta$ are equivalent (according to the theory of Kleene algebras)
when the Kripke structure associated with $\alpha$ satisfies exactly
the formulas associated with $\beta$. And this is indeed the case. Let
me make all of this precise.

\begin{figure}[t]
\begin{center}
\fbox{\begin{minipage}{6in}
We first construct the nondeterministic finite automaton with
$\epsilon$-moves $A^\epsilon_\alpha$, by induction on the structure of
$\alpha$:
\begin{align*}
A^\epsilon_a & \triangleq (\{q_0,q_1\},q_0,\{q_1\},\{(q_0,a,q_1)\})\\
A^\epsilon_{\alpha_1;\alpha_2} & \triangleq (Q'\dcup Q'',q'_0,Q''_f,\Delta'\dcup\Delta''\dcup (Q'_f\times\{\epsilon\}\times\{q''_0\}))\\
& \qquad\text{where $\begin{prog}
       A^\epsilon_{\alpha_1} = (Q',q'_0,Q'_f,\Delta')\\
       A^\epsilon_{\alpha_2} = (Q'',q''_0,Q''_f,\Delta'')\end{prog}$}\\
A^\epsilon_{\alpha_1+\alpha_2} & \triangleq (Q'\dcup Q''\dcup \{q_0\}, q_0, Q'_f\dcup Q''_f,\Delta'\dcup\Delta''\dcup (\{q_0\}\times\{\epsilon\}\times\{q'_0,q''_0\}))\\
& \qquad\text{where $\begin{prog}
       A^\epsilon_{\alpha_1} = (Q',q'_0,Q'_f,\Delta')\\
       A^\epsilon_{\alpha_2} = (Q'',q''_0,Q''_f,\Delta'')\end{prog}$}\\
A^\epsilon_{\alpha^*} & \triangleq (Q'\cup\{q_0\},q_0,Q'_f\dcup \{q_0\},\Delta'\dcup (Q'_f\times\{\epsilon\}\times\{q_0\}))\\
& \qquad\text{where $A^\epsilon_{\alpha} = (Q',q'_0,Q'_f,\Delta')$.}
\end{align*}
We derive the nondeterministic finite automaton without
$\epsilon$-moves $A_\alpha$ by identifying states that are reachable by
$\epsilon$-moves. Let $A^\epsilon_\alpha=(Q,q_0,Q_f,\Delta)$. Define
the following relations on states $Q$: let  $q\rightarrow^\epsilon q'$ if
there exists a sequence of states $q_1=q,q_2,\dots,q_{k-1},q_k=q'$ in
$Q$ such that $(q_i,\epsilon,q_{i+1})\in\Delta$, for $i\in 1..k-1$;
let $q\collapse q'$ if either $q\rightarrow^\epsilon q'$ or
$q'\rightarrow^\epsilon q$. The relation $\collapse$ is easily seen to
be an equivalence relation. Let $[q]_{\collapse}$ represent the
equivalence class of the state $q$, and let $[Q]_{\collapse}$
represent the set of equivalence classes $\{[q]_{\collapse}\mid q\in
Q\}$. The nondeterministic finite automaton $A_\alpha$ is obtained by
taking as states the $\collapse$-equivalence classes of states of
$A^\epsilon_\alpha$:
\[ A_\alpha \triangleq ([Q]_{\collapse},[q_0]_{\collapse},[Q_f]_{\collapse},
\{([q]_{\collapse},a,[q']_{\collapse})\mid (q,a,q')\in\Delta,
a\ne\epsilon\}).\]
\end{minipage}}
\end{center}
\caption{Construction of $A_\alpha$}
\label{f:construction}
\end{figure}

First, we need to construct a nondeterministic finite automaton
$A_\alpha$ corresponding to a program $\alpha$. There is nothing
original here. The trick is to do this in a way that clearly reflects
the structure of the program. (Otherwise, we end up pushing much of
the equivalence between programs into the construction of the
nondeterministic automaton, with the result of potentially begging the
question). In other words, we would like syntactically different
programs to yield different nondeterministic automata, even when those
programs are actually equivalent, such as $a$ and $a+a$. We can do
this most easily in two steps, first by inductively constructing a
nondeterministic finite automaton with $\epsilon$-moves (that is,
non-action moves that the automaton can perform at any time), second
by collapsing the automaton by identifying the states reachable by
$\epsilon$-moves, and  removing the $\epsilon$-moves. Recall that
a nondeterministic finite automaton is a tuple $A=(Q,q_0,Q_f,\Delta)$,
where $Q$ is the finite set of states, $q_0$ is the initial state,
$Q_f$ is a set of final (or accepting) states, and $\Delta$ is a set
of transitions, each transition being of the form $(q,a,q')$ and
representing a transition from state $q$ to state $q'$ upon action
$a\in\cA\cup\{\epsilon\}$. Figure~\ref{f:construction} summarizes the
construction of the nondeterministic finite automaton $A_\alpha$
corresponding to program $\alpha$.

To view a nondeterministic finite automaton as a Kripke structure is
 straightforward. The states of the automaton are the states of
the Kripke structure. We consider only the primitive propositions
$\Phi_0=\{\mi{init},\mi{final}\}$, where $\mi{init}$ says that a state is
initial, while $\mi{final}$ says that a state is final. The interpretation of
primitive propositions enforces this reading. The interpretation of
primitive programs is given by the transitions in the finite
automaton. Formally, if $\alpha$ is a program with
$A_\alpha=(Q,q_0,Q_f,\Delta)$, the Kripke structure
$M_\alpha=(S_\alpha,\pi_\alpha,\sigma_\alpha)$ corresponding to
$\alpha$ is given by:
\begin{align*}
 S_\alpha & \triangleq Q\\
 \pi_\alpha(q)(\mi{init}) & \triangleq
    \begin{cases}
    \true & \text{if $q=q_0$}\\
    \false & \text{if $q\ne q_0$}
    \end{cases}\\
 \pi_\alpha(q)(\mi{final}) & \triangleq
    \begin{cases}
    \true & \text{if $q\in Q_f$}\\
    \false & \text{if $q\not\in Q_f$}
    \end{cases}\\
 \sigma_\alpha(a) & \triangleq  \{(q,q')\mid (q,a,q')\in\Delta\}.
\end{align*}

We can now associate with a program $\beta$ (when viewed as a
specification) the set of formulas of PDL that $M_\beta$ satisfies,
$\intension{\beta}\triangleq\{\phi \mid M_\beta\sat\phi\}$. The following
result follows rather easily from Proposition~\ref{p:pdl-equivalence}:
\begin{proposition}
The following are equivalent:
\begin{enumerate}
\item[(a)] $\alpha=\beta$;
\item[(b)] For all $\phi$, $M_\alpha\sat\phi$ if and only if
$\phi\in\intension{\beta}$. 
\end{enumerate}
\end{proposition}
Of course, this statement is simply restating the well-known fact
that two regular expressions are equal (i.e., denote the same 
language) when their corresponding finite automata recognize the same
language.

Let me conclude this section with some comments on PDL. The logic we
have used is rather poor, in that it cannot be used to reason
about programs  with conditional statements. If we add to the
syntax of regular programs a class of programs of the form $\phi?$,
interpreted as ``if the current state satisfies $\phi$, then
continue'', we can encode a conditional such as
$\mathit{if}~\phi~\mathit{then}~\alpha_1~\mathit{else}~\alpha_2$ by
$(\phi?;\alpha_1)+((\neg\phi)?;\alpha_2)$. This addition makes the
syntax of programs and formulas in the logic mutually recursive. It
also means that we cannot give the semantics of programs independently
of the semantics of the formulas of the logic. With this in mind, we
extend $\sigma$ to tests with respect to a Kripke structure $M$ by
taking
\[ 
\sigma(\phi?)\triangleq \{ (s,s) \mid (M,s) \sat\phi\}. 
\]
If we restrict $\phi?$ to only use propositional formulas (without
occurrences of modal operators $[\alpha]\phi$ or $\<\alpha\>\phi$),
we get a logic sometimes called 
\emph{poor test PDL}. It is possible to reason equationally about
programs with poor tests by using a variant of Kleene algebras
called Kleene algebras with tests \cite{r:kozen97}. If we allow tests
$\phi?$ to use arbitrary PDL formulas, we get \emph{rich test
PDL}. Rich test PDL is very expressive; it lets us write formulas
that include programs such  as $[\alpha_1]\phi?;\alpha_2$, which says that if all
halting executions of $\alpha_1$ result in a state where $\phi$
holds, then execute $\alpha_2$.  It seems counterintuitive for programs to be
able to perform speculative execution in that way, especially since
such properties tend to be undecidable for reasonable programming
language. I know of no equational theory for programs using such
strong tests.

\section*{Conclusion}

As the examples above illustrate, there is often a deep relationship
between logical verification and equational verification. One might be
left with the impression that such relationships are always
present. Unfortunately, the more complex the equivalence, the more
difficult it is to capture through a logical specification. Some of
the most involved equivalences being applied nowadays occur in
cryptography, where a cryptographic scheme is generally proved correct
by showing it is equivalent to a simpler scheme which is
unimplementable, but more obviously correct (perhaps because it uses a
trusted third party, or a perfectly  private channel)
\cite{r:goldreich98}. A potentially
interesting venue for exploring logical characterizations of
equivalence for cryptographic schemes is the recent work of Datta et
al. \citeyear{r:datta04} that attempts to relate and unify such
equivalences with a notion of equivalence based on a stochastic
process calculus \cite{r:mitchell01}. 

Let me close on a remark prompted by my choice of examples. The astute
reader will have noticed that HML and PDL have much of the similar
flavor. They both use formulas involving actions and their
effects. There is a difference, however, in that formulas in PDL
describe the actions of a program in a particular environment (given
by a state of the corresponding Kripke structure) while formulas in
HML describe the actions of the environment on a particular process
(given by the process serving as a model). Thus, despite surface
similarities, the logics are meant to reason about quite different
things, somewhat dual to each other: processes as models and
environments in formulas, versus environments as models and programs
in formulas. I am curious of the extent to which this duality can be
made precise, and whether there are insights to be gained from it.

\end{document}